\documentstyle[12pt,moriond,epsf]{article}
\newcommand{\hi}{H\,{\sc i}}
\newcommand{\himf}{H{\sc i}MF}
\newcommand{\ahiss}{AH{\sc i}SS}
\newcommand{\mhi}{\mbox{$M_{\rm HI}$}}
\newcommand{\mhis}{\mbox{$M^*_{\rm HI}$}}
\newcommand{\msol}{\mbox{${\rm M}_\odot$}}
\newcommand{\phis}{\mbox{$\phi^*$}}

\newcommand{\rhohi}{\mbox{$\rho_{\rm HI}$}}
\newcommand{\og}{\mbox{$\Omega_{\rm g}$}}

\newcommand{\etal}{{\em et~al.\/}}

\newcommand{\kms}{\mbox{$\rm km\, s^{-1}$}}
\begin{document}
\heading{THE LOCAL H\,{\sc i} MASS FUNCTION}

\author{Martin Zwaan} 
{Kapteyn Astronomical Institute, Groningen, The Netherlands}
{$^{ }$ }

\begin{moriondabstract}
 The local \hi\ mass function (\himf), like the optical luminosity
function, is an important observational input into models of cosmology
and galaxy evolution.  It is a helpful framework for assessing the
number density of gas rich dwarf galaxies, which are easily missed in
optically selected galaxy samples, as well as for determining the
cosmological density of neutral gas at the present epoch.  For \hi\
masses larger than $10^8~\msol$ the \himf\ is determined with reasonable
accuracy and the same function is obtained from both optical and \hi\
selected samples of galaxies.  However, the faint tail below $\mhi <
10^7~\msol$ is still ill-constrained and leaves room for a population of
gas rich dwarfs or free floating \hi\ clouds which hypothetically could
contribute significantly to the local gas density.  Determining the
faint tail far below \hi\ masses of $10^7~\msol$ will be a great challenge
for the future. 
 \end{moriondabstract}

\section{Introduction}\label{why.sec}
 The \hi\ mass function (\himf), the \hi\ equivalent to the better known
optical luminosity function (LF) defines the number of galaxies per
cubic Mpc as a function of \hi\ mass \mhi.  This function is an
important constraint in theories of cosmology and galaxy evolution.  The
shape of the local \himf\ tells us how the neutral gas in the nearby
Universe is distributed over galaxies of different masses.  For example,
semi-analytical models of galaxy formation require detailed measurements
of the \himf\ and the LF to test the predictions (e.g., \cite{som98}). 
In contrast to the LF, the \himf\ should also constrain the number
density of gas rich dwarfs and low surface brightness (LSB) galaxies. 
It has been shown that these objects are easily missed in optical
surveys which are used to evaluate the LF \cite{dis76,mcg96},
but they could contribute a significant part of the \himf.  Also the
often speculated upon, but yet unidentified class of free floating \hi\
clouds are definitively missed in LFs but are hypothetical contributors
to the \himf. 

Besides the shape of the \himf, also the normalization is of great
importance to cosmology.  The integral over the function yields the
cosmological neutral gas density of the local Universe.  This
measurement anchors the estimates of the neutral gas densities at higher
redshift determined from damped Ly$\alpha$ systems, seen in absorption
in the spectra of background quasars.  These systems are generally
considered to be the high $z$ equivalents of the cold gas disks of
present day spiral galaxies.  The exact measurement of the gas density
as a function of redshift brackets theories of structure formation and
also helps in the understanding of the physical processes of gas cooling
and star formation.  An important question is whether the LSB galaxies
and the gas rich dwarfs make a significant addition to the cosmological
neutral gas density at the present epoch. 

\section{How To Measure an \himf}
 Different methods can be employed for evaluating an \himf.  The largest
variety lies in the sample of galaxies that is used to derive the
function.  Basically two classes of galaxy samples can be distinguished:
optically selected galaxy samples and \hi\ selected galaxy samples. This
section briefly describes both methods.

\subsection{\himf\ Based on Optically Selected Galaxies}
 An \himf\ can easily be derived from optical luminosity functions 
when a conversion factor from optical luminosity to \hi\ mass is used.  
Suppose the LF can be described by a Schechter function of the form 
\begin{equation}
\phi(L) dL = \phis\left(\frac{L}{L^*}\right)^\alpha \exp
-\left(\frac{L}{L^*}\right)d\left(\frac{L}{L^*}\right),
\end{equation}
where $L^*$ is the characteristic luminosity that defines the knee in
the LF,
\phis\ is the normalization, and $\alpha$ is the value that defines the
slope of the faint end. By using a relation between optical magnitude $M$
and \hi\ mass, for example 
 $\log \mhi = A - BM,$
an \himf\ can be derived:
\begin{equation}
\Theta (\mhi) d(\mhi) = 
\frac{0.4}{B}
 \,\phis\left(\frac{\mhi}{\mhis}\right)^{(\alpha+1)\frac{0.4}{B}-1}
 \exp -\left(\frac{\mhi}{\mhis}\right)^{\frac{0.4}{B}}
 d\left(\frac{\mhi}{\mhis}\right). 
\end{equation}
 The low-mass slope of the \himf\ is related to the faint-end slope
of the LF as $\alpha_{\rm HI}=(\alpha_{\rm opt}+1)\frac{0.4}{B}-1$.  Rao
\& Briggs \cite{rao93} applied this method by using different LFs and conversion
factors for different morphological types and added up the results to
obtain a global \mhi\ distribution function.  The \himf\ that they
derived is well-fit by a Schechter function with $\alpha \approx -1.1$. 
Intuitively one would believe that a trend of increasing \hi\ richness
for low luminosity galaxies ($B<0.4$) implies a steep low mass end of
the mass function.  However, Rao \& Briggs already stressed that for
realistic values of $B$ ($\sim 0.35$ \cite{fis75}),
$\alpha_{\rm HI} \approx \alpha_{\rm opt}$, especially for $\alpha_{\rm
opt}$ in the range $-1.0$ to $-1.3$. 

\begin{figure}[htb] 
\begin{centering}
\epsfysize=8.5cm
\epsfbox{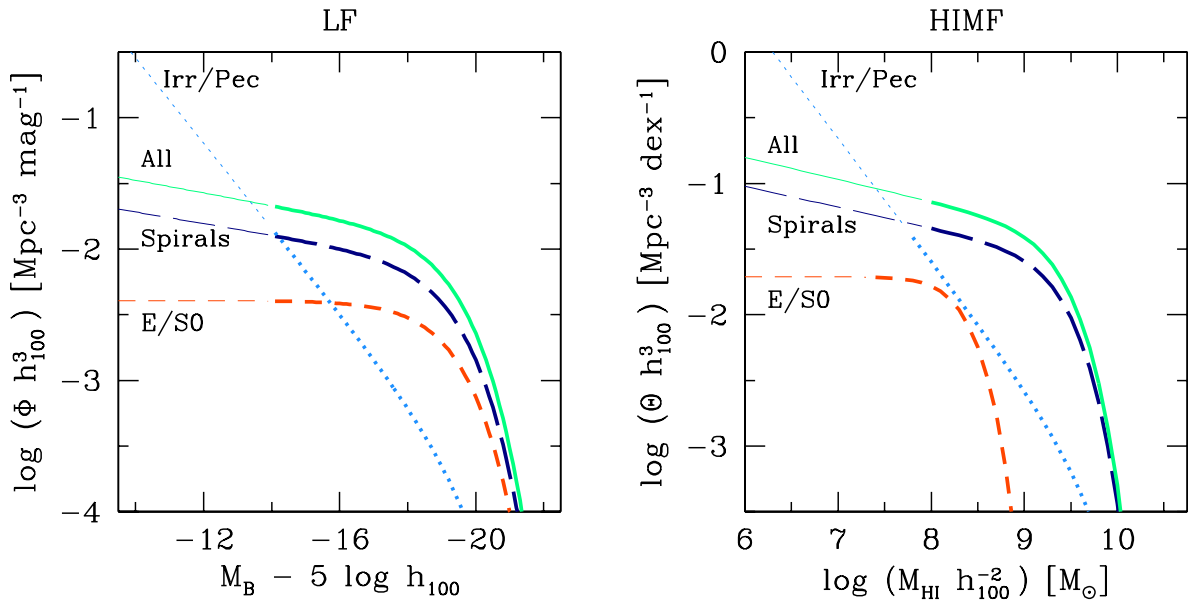}
 \caption{{\em Left panel:\/} Luminosity function for various
morphological types taken from Marzke \etal\ \cite{mar98}. {\em Right
panel:\/} converted \hi\ mass function following the procedure described
in the text.
 \label{marzke.fig}}
\end{centering}
\end{figure}

In Fig.~\ref{marzke.fig} an example of this method is shown.  The
luminosity function for various morphological types taken from Marzke
\etal\ \cite{mar98} is shown in the left panel.  The right panel shows
the derived \hi\ mass functions, using different values of $A$ and $B$
for the different morphologies.  It is clear that the \himf\ is
dominated by spiral galaxies, although at the very low mass end the
irregulars seem to overtake the number densities.  However, this
dominance occurs at luminosities below $M_B=-14$, the magnitude limit of
the Marzke \etal\ sample. 

A slightly modified form of this method of finding an \himf\ for
optically selected galaxies is to use the measured \hi\ fluxes for a
sample of galaxies with well-understood selection criteria.  For
example, Hoffman \etal\ \cite{hof89} conducted a series of pointed observations
with the Arecibo telescope of known galaxies in the Virgo cluster.  They
constructed a preliminary \himf\ and concluded that there is no excess of
gas rich dwarf galaxies.  Briggs \& Rao \cite{bri93} reanalyzed these data
supplemented with the Tully \& Fisher catalog of \hi\ observations of
spiral galaxies, and were able to construct the \himf\ over the range
$10^7$ to $10^{10}~\msol$.  They also concluded that there is no
evidence for a sharp rise in the number of gas rich dwarf galaxies, and
that in clusters the \himf\ might even go down. 

Although the estimation of the space density of gas rich objects using
galaxies that are selected on their star light is very instructive, it
does not yield an unbiased \himf\ for all gas containing objects.  The
\himf\ determined by optically selected galaxies defines a lower limit
to the ``real'' \himf, and it is not inconceivable that large
populations of extragalactic objects are missing from the analysis.  It
has become clear of late that low surface brightness (LSB) galaxies
contribute significantly to the number density of galaxies in the local
Universe (e.g., \cite{spr97,dal97}).  Since
these galaxies are found to be rich in HI \cite{sch92} they
are likely candidates to make a contribution to the \himf, while they
are easily missed in optical surveys \cite{dis76}.  Objects that are
most certainly missed in surveys in optical wavelengths are 
hypothetical free floating \hi\ clouds.  A conversion from a LF to a
\himf\ would never include these objects. 

\subsection{\himf\ Based on \hi\ Selected Galaxies}
 The most ideal method to measure an \himf\ is to use a ``blind'' survey
in \hi.  No biases against galaxies' surface brightnesses or magnitudes
are introduced that way.  The \himf\ can be constructed directly and
suffers no bias against LSB galaxies, gas rich dwarfs or even free
floating \hi\ clouds.  The major problem is that finding a galaxy in
\hi\ is much more difficult than finding it optically.  Blind surveys in
\hi\ take hundreds of hours of observing time to yield only few dozen
galaxies, while optical surveys systematically produce catalogs of
thousands of galaxies.  For illustration, the first published blind \hi\
survey in the field (Shostak \cite{sho77}) took many days of observing
time with the NRAO 300 ft telescope and yielded only one detection which
later turned out to be a high velocity cloud gravitationally bound to
the Milky Way galaxy.  In order to increase the detection efficiency
many surveys were pointed toward known overdensities: groups and
clusters of galaxies.  The M81, Sculptor, CVn I and NGC 1023 groups
\cite{hay79, lo79, fis81} as well as the the Hydra \cite{mcm93},
Hercules \cite{dic97}, Centaurus and Fornax \cite{bar97} clusters have
been surveys extensively in the 21cm line.  The conclusion that can be
drawn from these surveys are 1) the {\em shape} of the \himf\ is similar
to that that is derived in the field from optically selected galaxies,
2) no excess of gas rich dwarfs has been found, and 3) no \hi\ clouds
without stars are found.  Also global underdensities (voids) have been
surveyed \cite{szo94, kru84}, but no \hi\ selected galaxies have been
found there. 

A fair comparison of the \himf\ for \hi\ selected galaxies with that
from optically selected galaxies requires a blind \hi\ survey of the
field, with no preference to known over- or underdensities.  Henning
\cite{hen92, hen95} conducted a series of pointings on lines of constant
declination over a redshift range of 7200 km/s.  A total number of 39
significant detections were recorded, of which 50\% were previously
unknown.  While Hennings \himf\ seems to be indicative of an increasing
number of dwarf galaxies, the overall function lies below the lower
limit to the \himf\ set by counting the optically selected population. 

Two large surveys in the 21cm line have been conducted recently, both
with the Arecibo telescope.  The results from one of these surveys,
named \ahiss\ (Arecibo \hi\ Strip Survey) is the topic of the next
section.  The other survey, similar in size, is discussed in Spitzak
\cite{spi96} and Schneider \cite{sch98}.  There are currently surveys in
progress at Dwingeloo \cite{hen98}, at Arecibo (The Dual-Beam Survey
\cite{sch98}), and at Parkes \cite{kil98}.  The latter will cover the
whole southern sky out to 12,700 km/s.

\section{The Arecibo \hi\ Strip Survey}
 The survey was carried out during the period of August 1993 until
February 1994 when the Arecibo telescope was being upgraded.  Since the
pointing was immobilized, the data were taken in drift-scan mode.  Two
strips of constant declination ($\delta=14^\circ$ and $\delta=23^\circ$)
were traced, together covering 17 hours of RA.  The total sky coverage,
including the side lobes, was 65 square degrees, wile the survey depth
was $cz=7400~\kms$.  The minimal detectable \hi\ mass at the full depth
of the survey was $1.5 \times 10^8~h_{100}^{-2}~\msol$ in the main beam,
while $6 \times 10^5~h_{100}^{-2}~\msol$ of \hi\ could be detected at
$7~h_{100}^{-1}~\rm Mpc$. 

A total of 66 significant ($5\sigma$) detections were found, of which 32
are listed in existing catalogs of optically selected galaxies.  All
detections that are sufficiently far away from the Galactic equator
($b>10^\circ$) to avoid severe extinction by Galactic dust were imaged
in the $B$-band with the 2.5 Isaac Newton Telescope on La Palma.  {\em
All\/} \hi\ signals were identified in the optical images, and were
found to be ordinary galaxies, having both stars and gas.  No free
floating \hi\ clouds were detected in this survey, which puts serious
constraints on the existence of these objects which are potentially a
population of protogalaxies. 

21cm follow-up with the VLA in D-configuration was performed on all
significant detections.  These observations were required to determine
more accurate fluxes and positions for all sources.  Another important
reason for doing follow-up with higher angular resolution, is to see
whether the signals were caused by single clouds or galaxies, or by
pairs or small groups, whose line emission might stack up in the same
channels.  We found that this latter situation to occurs in five of the
61 cases.  For surveys which make use of dishes smaller than the Arecibo
telescope, and therefore scan the sky with larger beam sizes, this
effect is probably more important.  Higher resolution 21cm follow-up is
therefore essential to accurately determine the number of galaxies in
the survey, their \hi\ masses and to make adequate identifications with
optical images. 

From a comparison of the \hi\ and optical properties between the
cataloged and uncataloged galaxies in our \hi\ selected sample we
conclude that the uncataloged galaxies are preferentially the ones with
lower average optical surface brightness, lower luminosity, lower \hi\
mass, smaller linear sizes and higher gas to light ratios \cite{zwa98}. 
However, this does not imply that a blind \hi\ survey yields a
population of galaxies that has previously gone unnoticed in optical
galaxy surveys.  Simply because in these two strips the sky has been
surveyed deeper in \hi\ than in the optical, the uncataloged galaxies
will mostly be dwarf galaxies or galaxies with lower optical surface
brightness.  Furthermore, the restricted bandwidth of the receiving
system imposes a strict limit on the maximum redshift of the detected
galaxies, this limits does not exist in optical surveys.  These two
effects cause a predominance of low luminosity systems in an \hi\
selected galaxy sample.  It is well established that decreasing
luminosity and decreasing surface brightness correlate well with
increasing $M_{\rm HI}/L$ ratio \cite{van95, deb97}.  Therefore, the
differences found between cataloged and uncataloged galaxies are a
natural result of the survey technique and do not imply that the new
\hi\ selected galaxies have properties that set them apart from known,
optically selected galaxies. 
 
A \himf\ for this sample of \hi\ selected galaxies is determined by
using the $1/{\cal V}_{\rm max}$ method \cite{sch68}.  This method consists
of summing the reciprocals of the volumes corresponding to the maximum
distance at which each object could be placed and still remain within
the sample.  Advantages of this method are that the \himf\ is
automatically normalized and that it is nonparametric, that is it does
not use a Schechter function as an intrinsic assumption about the shape
of the \himf.  A recent overview of the different galaxy luminosity
function estimators is given by Willmer \cite{wil97}, who tests the validity
of different methods by means of Monte-Carlo simulations.  Careful
examination of his tables shows that the $1/{\cal V}_{\rm max}$ method (with
binning in magnitudes) recovers the input luminosity function
satisfactorily, and equally well as the more conventional parameterized
maximum likelihood method \cite{san79} or the
stepwise maximum likelihood method (SWLM \cite{efs88}). 

\begin{figure}[htb] 
\begin{centering}
\epsfysize=9cm
\epsfbox{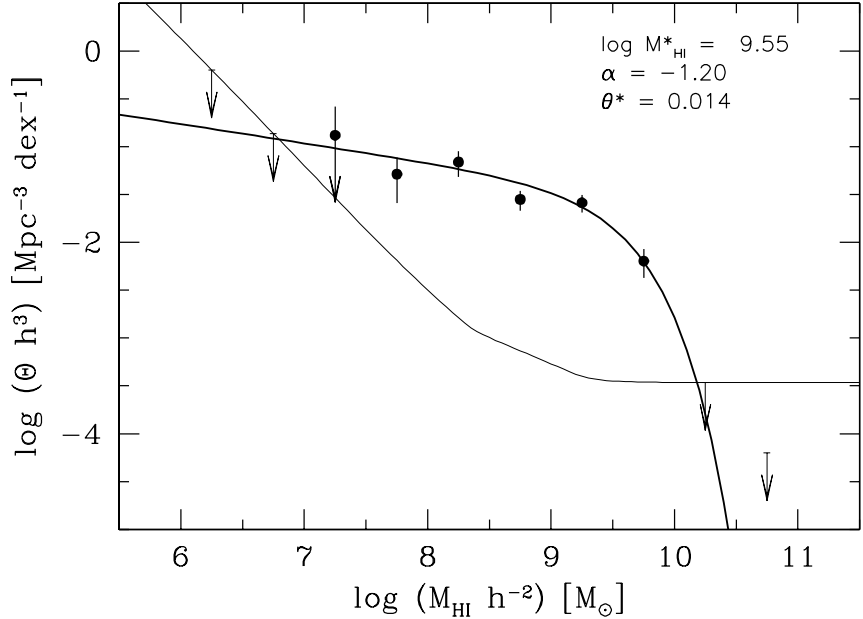}
 \caption{\hi\ mass function from the Arecibo \hi\ Strip Survey.  The
thin line is the sensitivity of the survey defined by $\phi=1/{\cal V}$,
where $\cal V$ is the effective search volume.  In the region $10^7 {\rm
M}_\odot < M_{\rm HI} < 10^{10} {\rm M}_\odot$ this function defines an
upper limit to the space density of intergalactic \hi\ clouds
without stars.  The measured \hi\ mass function per decade is shown
by the points.  The thick line is a Schechter luminosity function with
parameters as given in the upper right corner.  The arrows show upper
limits to the space density of \hi\ rich galaxies or intergalactic
\hi\ clouds.  The two arrows on the right are from a complementary
survey with the Arecibo telescope over the range 19,000 to 28,000 ${\rm
km\,s}^{-1}$. 
 \label{ahiss_himf.fig}}
\end{centering}
\end{figure}

Fig.~\ref{ahiss_himf.fig} shows the \himf\ for the \ahiss. The solid points
show the \himf\ per half decade, normalized such that they represent the
number of galaxies per $\rm Mpc^3$ per decade. Errorbars represent
$1\sigma$ errors from Poisson statistics. The solid line indicates a
best fit Schechter function to the points, parameterized by
 \begin{equation} \label{himf.eq}
 \Theta (\mhi) d\mhi
 = \theta^*\left(\frac{\mhi}{\mhis}\right)^{\alpha} \exp-\left(\frac{\mhi}{\mhis}\right) d\left(\frac{\mhi}{\mhis}\right),
 \end{equation}
 where $\alpha$ is the slope of the faint tail, $\theta^*$ is the
normalization, and $\mhis$ is the characteristic mass that defines the
knee in the \himf.  Because there are no \hi\ masses lower than $10^7$
or higher than $10^{10}~\msol$ detected by the \ahiss, the \himf\ is almost
unconstrained in these regions.  However, upper limits to the space
density of \hi\ rich galaxies or \hi\ clouds can be determined by
showing the sensitivity function in the same figure.  The sensitivity
function is defined by $\phi=1/{\cal V}$, where ${\cal V}$ is the effective search
volume.  The arrows indicate upper-limits to the \himf, on the left they
follow this sensitivity function, on the right they are determined by a
complementary survey with Arecibo over the redshift range $cz=$~19,000 to
28,000 \kms.  The best fit parameters of the Schechter function are
$\alpha = -1.20$, $\theta^*=0.014~ \rm Mpc^{-3}$ and $\log (\mhis/\msol)
= 9.55$. 

\subsection{The integral \hi\ density}
 As was mentioned in section~\ref{why.sec}, the local \himf\ can be used
to determine the integral gas density, \og, at $z=0$.  At high redshifts
\og\ can be derived from the study of damped Ly$\alpha$ systems, high
column density gas clouds seen in absorption in the spectra of
background quasars.  At low redshifts, these measurements are hampered
by poor statistics for several reasons: due to the expansion of the
Universe the expected number of absorbers decreases with decreasing
redshift, the Ly$\alpha$ line is not observable from the ground for
redshifts smaller than 1.65, and starlight and dust in the absorbing
foreground galaxies hinder the identification of the background quasars. 
The most recent determinations of the gas density from damped Ly$\alpha$
systems show that the redshift range from 0 to 1.65, which corresponds
to more than three quarters of the age of the Universe ($q_0=0.5$), is
covered by only about 20 systems \cite{tur97}.  Furthermore, the
lowest redshift damped Ly$\alpha$ system currently known is at $z=0.09$
\cite{rao98}, illustrating that a $z=0$ point can not yet reliably be
derived from damped Ly$\alpha$ observations. 
 
The gas density detected in the \ahiss\ can be readily determined by
taking the integral over the \himf\ multiplied by \mhi.  This yields
$\rhohi=\Gamma(2+\alpha) \theta^* \mhis$, where \rhohi\ is the neutral
gas density in $\msol/\rm Mpc^3$.  Using the values from
Fig.~\ref{ahiss_himf.fig} we derive that $\rhohi=5.8 \times
10^7~h_{100}~\msol / {\rm Mpc}^3$, corresponding to $\og=(2.7 \pm 0.5)
\times 10^{-4}~h_{100}^{-1}$.  This value corresponds very well with
earlier estimates based on optically selected galaxies by Rao \& Briggs
\cite{rao93}, who found $\rhohi=4.8 \times 10^7~h_{100}~\msol / {\rm
Mpc}^3$.  This value has been confirmed by Natarajan \& Pettini
\cite{nat97} who have used LFs from recent large redshifts surveys. 
Also the first results of the H{\sc i}PASS survey \cite{kil98} yield a
similar value.  When using their Schechter values $\alpha = -1.35$,
$\theta^*=0.014~ \rm Mpc^{-3}$ and $\log (\mhis/\msol) = 9.50$, we
derive $\rhohi=6.1 \times 10^7~h_{100}~\msol / {\rm Mpc}^3$. 

\begin{figure}[htb] 
\epsfysize=8.5cm
\epsfbox{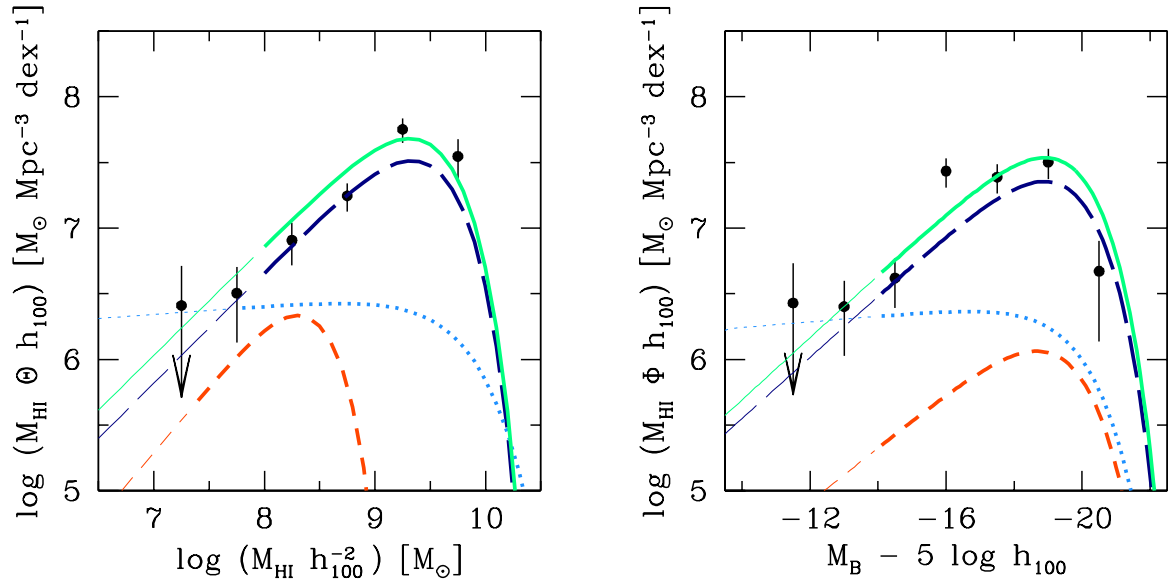}
 \caption{{\em Left panel:\/} Space density of \hi\ mass per decade
contained in objects of different mass.  {\em Right panel:\/} Space
density of \hi\ mass per mag contained in objects of different absolute
magnitude. Lines are as in Fig.~\ref{marzke.fig}. 
Most of the \hi\ at the present
epoch is locked up in luminous galaxies with high \hi\ masses. 
Dwarfs seem to contribute very little to the gas density. 
 \label{dens.fig}} \end{figure}

Figure~\ref{dens.fig} shows how this gas density is distributed over
different \hi\ masses and different luminosities.  As in
Fig.~\ref{marzke.fig} the lines represent converted LFs \cite{mar98} for
different morphological types.  It is clear that the gas density in
the local Universe is dominated by massive galaxies in the range
$10^9~\msol < \mhi < 10^{10}~\msol$, with high luminosities.  The
presently known dwarf galaxies are minor contributors to \og\ since
galaxies with \hi\ masses below $10^8~\msol$ are responsible for only
$\sim 20\%$ of the gas density.

Although the run of \og\ with redshift is generally considered to reach
a peak at $z\approx 3$ and then declines until $z=0$, the most recent
(preliminary) results from Turnshek \etal\ \cite{tur97} show that $\og
(z)$ is consistent with no evolution from $z=3$ to $z\approx 0.15$. 
Still, the $z=0$ point, evaluated via the 21cm in emission, lies
approximately a factor of six lower.  If, like the \ahiss\ results
suggest, most of the neutral gas indeed resides in large spirals, the
identification of galaxies responsible for damped Ly$\alpha$ absorption
would be trivial.  Rao \etal\ \cite{rao98} show that the opposite is
true: none of the galaxies in the vicinity of the quasar in which two
nearby absorbers are found are luminous spirals.  Also at higher
redshifts there are indications that dwarf galaxies contribute
significantly to the cross section of damped Ly$\alpha$ absorbers.  Le
Brun \etal\ \cite{leb97} find that most of the optical counterparts of
damped Ly$\alpha$ systems at high redshift are dwarf galaxies. 
Matteucci \etal\ \cite{mat98} conclude on the basis of chemical
evolution models that most damped Ly$\alpha$ absorbers are in fact dwarf
galaxies. 

There are two possible solutions to this $z=0$ conundrum.  Either there
has been strong evolution at very low redshift, implying that most of
the star formation has occurred over the last few Gyrs, or estimates of
the gas density at $z=0$ from the 21cm line are still incomplete.  At
least in the range $\mhi>10^8 ~\msol$ most authors agree well on the
shape and normalization of the \himf\ and derivations based on optically
selected and \hi\ selected galaxies are consistent.  If an
incompleteness of the \himf\ exists it must be in the tail of the \himf. 
The question to answer is therefore: are there large populations of gas
rich dwarf galaxies or \hi\ clouds with typical 21cm fluxes undetectable
in present surveys?

In this light it is interesting to note that are recent claims of a
sharply rising LF, typically faintward of $M=-16$ (see \cite{lov98}). 
The galaxies responsible for this upturn are generally of late
morphological type, with blue colors and low optical surface brightness. 
These are the galaxies normally regarded as being rich in neutral gas
(see also \cite{sch98}). 

Another interesting recent result comes from Blitz \etal\ \cite{bli98},
who argue that many of the high velocity clouds (HVCs) populated around
the Milky Way Galaxy are actually \hi\ clouds with masses of a few times
$10^7~\msol$ distributed trough the Local Group.  Their integrated mass
may account to as much as $10^{11}~\msol$. 

\section{The Faint Tail}
 Obviously, determining the faint tail of the \himf\ deserves special
attention.  An accurate determination below $\mhi=10^7~\msol$ has proven
to be an extremely difficult exercise in blind surveys.  For the Parkes
Multibeam survey, the largest \hi\ survey to date, the limiting depth
for a galaxy with $10^6~\msol$ of \hi\ is 1 Mpc \cite{sta96}.  This
estimate assumes that the profile width of the galaxy is 200 \kms, while
in reality the profile width for these tiny \hi\ masses will probably be
much smaller, implying that the same flux is spread over fewer channels
which makes the galaxy more easily detectable.  Even if the limiting
depth is 2 Mpc, the maximum search volume for $10^6~\msol$ \hi\ masses
is ${\cal V}=\frac{1}{3}\Omega\, D^3 \approx 17\, \rm Mpc^3$, where
$\Omega$ is the solid angle.  The search volume will therefore be
confined to the Local Group environment. 

For a blind \hi\ survey, the survey volume as a function of \hi\ mass
can be evaluated in the following way.  The minimal detectable \hi\ mass
$\mhi_{\rm ,min}$ for a $5\sigma$ detection can be expressed as $236
\times 5\sigma \Delta V D^2$, where $\sigma$ is the noise in the
detection spectra in mJy, $\Delta V$ is the profile width in km/s and $D$ is the
distance to the source in Mpc.  Assuming optimal smoothing of the survey
spectra, the noise goes as $\sigma \propto \Delta V^{-1/2}$, and therefore
$\mhi_{\rm ,min} \propto D^2 \Delta V^{1/2}$.  The limiting depth, the
maximum distance at which a source could be placed and still be
detected, is then given by $D_{\rm c} \propto \mhi^{1/2} \Delta V^{-1/4}$. 
The \hi\ mass is related to the profile width via the Tully-Fisher
relation and the relation between \hi\ mass and optical luminosity.  The
``\hi\ Tully-Fisher relation'' goes as $\Delta V \propto \mhi^{1/3}$
\cite{bri93}.  The limiting depth is therefore  $D_{\rm c} \propto
\mhi^{5/12}$.  Finally, since the total search volume is
${\cal V}=\frac{1}{3}\Omega\, D_{\rm c}^3$, we derive ${\cal V} \propto \mhi^{5/4}$. 
In contrast to optical surveys, \hi\ surveys are conducted with a
limited bandwidth which imposes a hard upper limit to the limiting depth
for massive galaxies.  The surveys volume therefore reaches a maximum
for \hi\ masses that can be detected at the redshift corresponding to
the receiver's frequency limit and does not increase for higher masses. 
The sensitivity curve, the inverse of the survey volume, is therefore a
straight line with slope $-1.25$, leveling off to a straight line for
\hi\ masses which can be seen all trough the survey volume.  The
decreasing part is parallel to an \himf\ with a slope of $-2.25$. 

The number of detections in a survey can be estimated by taking the
integral over the \himf\ times the survey volume: 
 $$
 {\cal N}=\int_{0}^{\infty}\,\Theta(\mhi) {\cal V}(\mhi)\, d\mhi.
 $$ 
 Graphically, this number can be regarded as the surface enclosed by the
\himf\ and the sensitivity curve (see Fig.~\ref{ahiss_himf.fig}).  If
$\alpha<-2.25$ the surface would not be closed and the number of
detections would be infinite.  For the {\sc HiPASS} survey, the
instrumental depth limit set by the bandwidth of the receiving system is
$127~h^{-1}_{100}~\rm Mpc$ and the covered solid angle is $\Omega=6.5$. 
If then the Schechter parameters for the \himf\ found by Zwaan \etal\
\cite{zwa97} are used, the number of detections for this survey will be
${\cal N}=2516$ galaxies, increasing the size of existing samples by a
factor of 30.  The statistics will be dominated by \mhis\ galaxies: more
than 70\% of the detected galaxies will have \hi\ masses in the range
$10^9~\msol < \mhi < 10^{10}~\msol$.  If the faint slope of
$\alpha=-1.2$ extends to the lowest masses, only $4 \pm 2$ galaxies with \hi\
masses below $10^7~\msol$ will be detected.  This number rises to 13 if
the latest value of $\alpha=-1.35$ quoted by Kilborn \cite{kil98} is
used.  If the slope is much steeper than currently assumed, for
example $\alpha=-1.7$ below $\mhi=10^8~\msol$, in concordance with the
estimates of the optical LF, the number of dwarfs will be 57, high
enough to reliably determine the shape of the \himf\ down to
$\mhi=10^6~\msol$. 
 
Thanks to their smaller dishes, synthesis instruments like the WSRT and
the VLA sample larger volumes per single pointing than the Parkes
telescope.  These instruments may be more useful for constraining the
faint end.  If the accessible bandwidth is 20 MHz the sampled volume
will be ${\cal V} = \frac{1}{3} D_{\rm max}^3 ({\rm FWHM}/2)^2\pi = 1.5~\rm
Mpc^3$ per single pointing.  In 24 hours \hi\ masses of $10^7~\msol$
will be detectable throughout the volume.  Adopting the \himf\ parameters
for field galaxies from the \ahiss\ we estimate that the number of
detections will be approximately half a galaxy per pointing of 24 hours. 
The statistics will be dominated by galaxies in the range $10^7~\msol <
\mhi < 10^8~\msol$, two orders of magnitude lower in \hi\ mass than in
the {\sc HiPASS} survey.  If again the \himf\ with $\alpha=-1.7$ for
$\mhi < 10^8~\msol$ is assumed, the number of detections in the $6$ to
$6.5~\msol$ bin will be 0.05.  In order to get a statistically
significant result many months of observing time will be
required.  Apparently, determining the faint end slope of the \himf\
below $10^7~\msol$ in the field is not possible with the current
generation of radio telescopes. 

A way to overcome the poor statistics and the uncertain volume
corrections is to study nearby groups of galaxies.  Typical
overdensities of these groups are a factor 25 of the cosmic mean, which
improves the statistics significantly.  Furthermore, the distances of
groups can be determined quite accurately using Cepheid measurements or
by photometry on the brightest resolved stars of only one or a few group
members, or by using the {\sc potent} method \cite{ber90}.  Studies of
the HI content of galaxies in different environments have shown that the
{\em shape} of the \himf\ for $\mhi > 10^8$ is independent of cosmic
density.  This finding justifies the study of the \himf\ in groups of
galaxies, although it is possible that the faint tail may depend on
environment. 

\acknowledgements{
Thanks to F. Briggs, R. Swaters and E. de Blok for comments on the text.
Financial support was received from the EC (TMR) and
the LKBF.}


%
\vfill
\end{document}